\documentclass[sigconf,nonacm,10pt]{acmart}
\setcopyright{none}
\acmDOI{}
\acmISBN{}

\usepackage{alltt}

\begin{document}

\title{Making Software Meaningful}

\author{Eagon Meng}
\email{eagon@mit.edu}
\affiliation{
  \institution{MIT}
  \city{Cambridge}
  \country{USA}
}

\author{Abutalib Namazov}
\email{abutalib@mit.edu}
\affiliation{
  \institution{MIT}
  \city{Cambridge}
  \country{USA}
}

\author{Carmel Schare}
\email{schare@mit.edu}

\affiliation{
  \institution{MIT}
  \city{Cambridge}
  \country{USA}
}

\author{Alcino Cunha}
\email{alcino@di.uminho.pt}
\affiliation{
  \institution{University of Minho \& INESC TEC}
  \city{Braga}
  \country{Portugal}
}

\author{Daniel Jackson}
\email{dnj@mit.edu}
\affiliation{
  \institution{MIT}
  \city{Cambridge}
  \country{USA}
}


\begin{abstract}
Adopting a single measure can improve the usability, modularity and accountability of software: a commitment to \textit{explicit meaning}. This entails constructing and agreeing upon a representation of the behavior of the software, as observed in the domain of application. The phenomena comprising this behavior become a vocabulary that grounds all discourse about the software, among all stakeholders, and for all artifacts and activities.

These phenomena are \textit{individuals}; \textit{actions} they participate in; and \textit{facts} that result from actions. They can be organized, by partitioning the set of actions, into \textit{concepts}, offering larger units of meaning.

Examples of exploiting meaning are given in three areas: designing for usability (by aligning user and designer on a single shared meaning); generating modular code with LLMs (by mapping units of meaning to units of code, achieving not only modularity but also legibility); and making agents accountable (by having them adhere to a code of conduct that defines their intended behavior).

\end{abstract}
\maketitle

\section{Introduction}

Many problems of software, in both its use and development, arise because there is no clear and ubiquitous interpretation of its behavior. What is missing is a \textit{meaning} that is shared between all stakeholders---users, designers, engineers, and more---and applied consistently across all activities and artifacts.

Take, for example, the confusion experienced by users when Facebook added new ``reactions'', including one showing an angry face. Users understood that liking a post was an \textit{upvote}, making the post more likely to be promoted and appear in newsfeeds. But was clicking the angry button an upvote? Or was it perhaps a downvote?

Every programmer has tried to trace a bug observed in the user interface down through the code, only to discover that what seemed to be a simple action in the interface did not map to a simple function in the code. Instead, they find themselves on a brambly path of asynchronous calls, routing and indirections, never certain of which part of the code was responsible for the observed effects.

A performance engineer trying to track down the source of a latency problem, or an analytics engineer trying to determine how much a new feature is used, might seek an answer in an event log. But they are likely to find that the events recorded in the log do not correspond in a straightforward way to actions performed by users, so elaborate queries must be formulated to massage the log into a form that conveys an intuitive picture of what is actually going on.

All these cases involve the occurrence of actions that carry meaning in the mind of the user and in the domain of application. And yet those actions and their meaning are lost in the user interface, in the code, and in the event log. This happens primarily because there is no \textit{shared understanding} of that meaning across artifacts, activities and roles.

This paper proposes a remedy that is simple in principle, and, we believe, tractable (if a bit more challenging) in practice. It is to \textit{make software meaningful}, by constructing a lightweight ontology of actions, facts and individuals through which the behavior of the software can be compellingly explained, and which can be applied in all aspects of development by all participants.

The ingredients of this idea are not new. We regard this as a merit rather than a flaw, because familiarity with the basic elements should make the approach easier to use and more likely to be adopted. The idea itself is, to our knowledge, a novel contribution, even if aspects of the idea have been articulated many times before.

We start with a brief history of the intellectual currents in computer science concerned with giving meaning to programs (Section \ref{history-section}). We then motivate the importance of meaning by explaining, in brief, the role that it plays in achieving usability, modularity and accountability (Section \ref{why-section}). A structure for expressing meaning is then described (Section \ref{structure-section}). Three applications are then presented that show the benefits of aligning meaning: in usability design (Section \ref{usability-section}), in generating code with LLMs (Section \ref{code-gen-section}) and in designing agents for autoresearch (Section \ref{agent-section}).

\section{A Brief History of Meaning}\label{history-section}

Efforts to attribute meaning to software go back to the earliest days of computer science. Debates about the role and significance of software, and more technical arguments about what notions such as ``correctness'' entail, are often, at their core, actually about \textit{meaning}: how software is mapped in the minds of its users, designers and implementers into a form that transcends the details of its representation on screen or in code.

\subsection{Traditional Language Semantics}
The field of programming language semantics established early on that the meaning of a program can be found in its observable behavior. All forms of semantics---denotational, operational, axiomatic---draw on observations of inputs and outputs, states and steps, even as they differed in what exactly those observations comprised, in their granularity, and in how they were obtained from the program text. The catchy title of Jean Raymond Abrial's book \cite{abrial-book} on deriving programs from B specifications, ``Assigning Programs to Meanings,'' marked an important shift (which had actually occurred long before): that a program's meaning could exist independently (and prior to) the program itself.

\subsection{Finding Meaning in the Domain of Application}
While language theorists viewed semantics in terms of the physical computer (or in terms of a mathematical abstraction thereof), software engineering researchers---and others who were both more practically and more philosophically inclined---recognized that the true meaning of a program must reside in its domain of application. In response to Edsger Dijkstra's assertion that ``the only thing a computer can do for us is to manipulate symbols,'' Terry Winograd responded: ``In my vicinity of the ‘real world’, I see computers doing lots of other things. They issue payroll checks, control the motion of metalworking machines, format and print architectural drawings and newsletters, and keep my car’s brakes from locking. They may manipulate symbols (and also manipulate electrical and magnetic fields), but that is instrumental—the means to an end.''~\cite{dijkstra-winograd}.

Dijkstra's conviction that software is, in essence, symbolic, embodies an amusing contradiction. What makes a token or mark a \textit{symbol} is precisely that it represents or points to something beyond itself. Brian Cantwell Smith explored this idea deeply, and lamented the lack of ``genuine theories of content'' which would address ``what it is that makes a given symbol or structure or patch of the world be about or oriented towards some other entity or structure or patch.''
\cite{cantwell-smith}
\footnote{Perhaps Computational Reflections \cite{cantwell-smith-computational-reflections-book}, a posthumous volume just appearing, was intended precisely to fill this void. Brian will be remembered admiringly for his insightful contributions to the philosophy of computing, and his early death is a tragic loss to our community.} Michael Jackson locates the meaning of software in its interaction with the world, in terms of phenomena that are shared between domains: on the one hand, a software machine, and on the other, the world in which the machine operates (which encompasses its users) \cite{jackson-world-machine}.\footnote{Jackson would actually identify, for most systems, not just two domains corresponding to the machine and the environment, but a network of domains that includes ``connection domains'' to account for the gap between the specification (about prescribed behavior at the interface between world and machine) and the requirement (about intended effect in the world) \cite{jackson-problem-frames-book}.}

\subsection{Formal Methods}

Work in the field of formal methods has been pursued in two threads, sometimes intertwined but often separate. One has focused on the verification of code; the other on languages for modeling behavior. The communities and tools of these threads have often overlapped. Verification, after all, needs a specification to define correctness, and modeling needs verification of model properties.

Nevertheless, a cultural gap emerged between, on the one hand, work on languages like Alloy, B, VDM and Z, which sought to capture the essential behavior of software systems, and, on the other hand, work on strategies and tools for verifying code. Often it seemed as if the two never met: models were not related to code, and verification was limited to low-level properties or to infrastructural components like file systems and compilers\footnote{There have been a few notable exceptions \cite{torlak-radiotherapy, abrial-paris-railway}, but full-blown verification of code has turned out to be so burdensome that moving even part of the way towards observable behavior requires heroic efforts. Some researchers look to LLMs as saviors, hoping that they will be able to synthesize loop invariants and function specifications automatically \cite{bertrand-meyer-ai-verification}.}.

As verification became a more popular area of research, interest in system-level modeling within the formal methods community waned, and with it serious attention to the question of what software \textit{means} in the context of its domain of application.

\subsection{Conceptual Modeling}

Usability experts have for decades pointed to the centrality of the ``conceptual model'' in providing a shared understanding of meaning between users and designers \cite{don-norman-book}. And yet the discipline of human-computer interaction seems to have never addressed the question of what form this model should take\footnote{Researchers at PARC in the 1970s and 80s explored how people formed mental models of mechanisms \cite{dekleer1981mental}, but they did not pursue the practical question of how to represent let alone design a conceptual model.}.

One might have hoped that the field of ``conceptual modeling'' would have provided an answer. But that field has not been motivated to produce the kind of conceptual model needed for usability, and has tended to be more interested in representations of \textit{data} (such as knowledge graphs and ontologies) than in representations of \textit{behavior}. Much valuable work has been done to produce languages and logics that allow subtle distinctions to be made. But the challenge---more important in our view---of giving meaning to software has fallen by the wayside.

\subsection{Concept Design}

Concept design attempts precisely to fill this gap, and to provide a simple and practical way to represent the meaning of a software system, arguing that the most consequential aspect of software design is the design of this meaning \cite{jackson2021essence}. 

No new elements are needed to construct such a meaning; concept design found them all hiding in plain sight. From formal methods, it took the idea of modeling behavior as a state machine consisting of actions over a (typically infinite) set of states. From conceptual modeling, it took the idea of representing entities in the world (including users) as atomic individuals with persistent identity, and the idea of representing state as a collection of facts about these individuals and their relationships to one another.

Concept design sought a notion of ``concept'' that not only aligned with the intuitive interpretation that users of software give to the word---the ``concept of layers in Photoshop'', the ``concept of friending in Facebook''---but that also had a simple interpretation in terms of these phenomena of states, actions and individuals. The solution was to think of a software system as a collection of state machines executing independently (and coordinated when need be), with each concept associated with a single machine. In architectural terms, a concept is thus similar to a microservice (although more truly ``micro'', since a conventional microservice would comprise many concepts).

Put simply, structuring behavior into concepts involves partitioning the actions: each action goes into a single concept or module, along with its associated state components. The resulting concepts correspond to activities or functional concerns; they are more like verbs than nouns. In contrast, object-oriented design treats the (types of) individuals as modules, which has an elegant parsimony but makes it hard to separate concerns. In a concept design, a user might have a user name and password in one concept (for Authenticating) and a display name and bio in another (for UserDisplaying, say). An object-oriented design would by default conflate the separate concerns of authenticating and user displaying, and define all these state components as attributes of the same object.

\subsection{From Concepts to Meaning}

Initial work on concept design \cite{jackson2021essence} was focused on the idea of concepts as reusable patterns, and on the impact that the design of individual concepts has on usability.

As concept design has been applied in other areas, though, notably in structuring implementations, generating code with LLMs, and most recently governing the behavior of AI agents, a more basic idea has emerged. That idea is the subject of this paper, and is, in short, that much can be gained from defining the essential behavior of a software system in a meaning that is ubiquitous and applied consistently across all phases of development, and shared between all roles in an organization. Thus the same actions presented to users are API endpoints in the underlying services and events in the execution log.

This idea might seem obvious, and indeed it has roots in the earliest work in formal specification. Good programmers frequently seek this kind of alignment, taking terms from the problem domain as names for functions and events. But to our knowledge, this idea has never been pursued systematically.


\section{Why Meaning Matters}\label{why-section}

Making software meaningful is not an end in itself, but is a prerequisite for other desirable properties. Consider three such properties---usability, modularity and accountability---which will be illustrated more fully in later sections.

\subsection{Usability}
Usability is largely about meaning. Usable software makes its meaning clear, and aligned with user goals. A user can then easily translate intent into action, and can interpret the results that follow. 

In contrast, there may be a \textit{gulf} between user and software: a gulf of execution that prevents users from mapping intents to actions, and a gulf of evaluation that prevents them from interpreting the results. This idea, originally proposed by Ed Hutchins, Jim Hollan, and Don Norman, has been helpful in pinpointing the barriers to usability of poor user interfaces (and was formulated in large part to explain the virtues of direct manipulation)~\cite{Hutchins1985DirectManipulation}. It offers a useful metaphor also for usability problems that arise from deeper problems of missing or mismatched meaning. When there is no common meaning shared by the user and the software, the gulf may be unbridgeable (and 
no amount of interface tweaking will help).\footnote{It should be noted also that the gulf metaphor assumes an execution model of a task that is fulfilled in a single step. The problems that arise from unclear or misaligned meaning often involve multiple steps, with the user struggling to identify the sequence of steps that will lead to the eventual, intended goal.}

\subsection{Modularity}
Modularity in its essence is also about meaning. A codebase is not modular just because it is divided into distinct components with well-defined interfaces. It must be organized in a meaningful way, with the components representing aspects of the functionality that have a coherent meaning in their own right. If so, the structure of the software's function (that is, its specification) and the structure of its code (that is, the implementation) will correspond, and all code-related tasks, from generating the code in the first place to adding new features, will be eased. In particular, if orthogonal pieces of functionality are mapped to separate modules, it may be possible to implement the modules entirely independently of one another. This has major implications for LLM-based software development, since it makes the generation of code simpler and less fragile, and reduces the context required (and thus the cost too).

\subsection{Accountability}
Accountability is ensuring that the behavior of software is apparent to its stakeholders and that it conforms to their expectations. It therefore is also about meaning: demanding a meaning that is effectively and honestly conveyed to stakeholders, and bounded appropriately. Accountability is a concern for traditional software, as evidenced by privacy regulations such as the GDPR. In a world of AI agents that have access to their owners' confidential information, and that can perform critical tasks on their behalf, the need for accountability becomes even more critical. If agentic software is made to be meaningful, it will be possible to understand and control what it is doing.

\subsection{Meaning and Shared Language}

In all three cases, meaning serves to create a sharing of  understanding. For usability, the sharing is between users and designers, and the shared meaning is the conceptual model---which must be at the same time the design model for the designer and the mental model for the user. For modularity, the sharing is between requirements analysts and programmers, who must find a structure for describing functionality that will serve both for planning features and for organizing code. For accountability, the sharing is between the agents themselves (or their developers, to the extent that they can predict their behavior), the users that deploy them and the regulators that oversee their usage.

Many problems in software development arise from the lack of a shared language---between teams and between roles. It is common for programmers to extensively rework functionality that was specified by product managers or user experience architects as they encounter difficulties during implementation. Such problems arise in large part because they speak different languages: journey maps and personas on the one hand, and data models on the other. The ubiquitous language of domain driven design \cite{evans-domain-driven-design} is of great value because it encourages an agreement on a vocabulary that is shared by an entire team. Sharing meaning takes this idea one step further, beyond a shared vocabulary to a shared interpretation of phenomena and their relationships to one another.

\section{A Structure for Meaning}\label{structure-section}

If a software system has a meaning, what form should that meaning take, and how can it best be expressed? The question of expression is about syntax and, while crucially important in practice, is secondary to the question of form, which will be the focus here.

\subsection{An Ontology of Phenomena}
The elements that comprise the form are drawn from a simple ontology of phenomena that is grounded in long established software engineering practice. There are four kinds of phenomena:
\begin{itemize}
\item \textbf{Individuals}. An individual represents a unique entity with a persistent identity. Individuals can have limited or unlimited lifetimes, and may come into existence at some time and disappear at another.
\item \textbf{Values}. A value represents something that can be interpreted\footnote{An individual---unlike a value---cannot be \textit{interpreted} and has no structure. It has its identity alone, and identities can be matched but not compared or decomposed in any other way. The conventional distinction between ``entity'' and ``value'' objects implies a setting in which both are composite structures. It is essential for our theory that individuals are \textit{not} composite, and do not ``contain'' attributes, but rather are associated with values and other individuals through facts. This relational view is critical to escaping from the conflation that object orientation produces, preventing effective separation of concerns \cite{jackson-beyond-objects}.} by virtue of its structure or by comparison to other values. Thus dollar amounts, phone numbers and email addresses are values; so are mailing addresses, domain names and the contents of messages.
\item \textbf{Actions}. An action is an atomic occurrence that corresponds to a meaningful happening, and involves some participants that are individuals\footnote{The actors that perform actions are usually represented as individuals, but there are other individuals that represent inert entities. Some actions are initiated by individuals; others are performed by the system. The participating individuals of an action may include the action's initiator, but this is not necessary: an action can be performed anonymously.} and values. The participants are classified into inputs and outputs. An action can have any number of outputs.
\item \textbf{Facts}. A fact is an assertion about a single individual, or about a relationship between individuals (or between an individual and a value).
\end{itemize}

As an example, an authenticating behavior may include:
\begin{itemize}
\item Individuals {\tt\small user1}, {\tt\small user2} and {\tt\small user3} corresponding to the users;
\item Values {\tt\small `alice'}, {\tt\small `bob'} and {\tt\small `carol'} corresponding to user names, and other values for their passwords;
\item Actions like {\tt\small register (`alice', `foo'): (user1)} in which Alice registers with the user name {\tt\small `alice'} and the password {\tt\small `foo'}, creating a user {\tt\small user1};
\item Facts such as {\tt\small registered(user1)} which says that {\tt\small user1} is a registered user, and {\tt\small username(user1, `alice')} which says that the username of {\tt\small user1} is {\tt\small `alice'}.
\end{itemize}

In general, actions may be contingent on certain facts holding, and their effect is to add new facts and remove old ones. Thus the action {\tt\small login (`alice', `foo'): (user1)} is contingent on {\tt\small username(user1, `alice')} and {\tt\small password(user1, `foo')}. These facts would be added by the action {\tt\small register (`alice', `foo'): (user1)}, which  itself is contingent on other facts, in particular that {\tt\small username(u, `alice')} is not true for any user {\tt\small u}, and that {\tt\small registered(user1)} is false.

\subsection{Defining Meaning}
To define the meaning of a software system, we do not, of course, enumerate all individuals, actions and facts explicitly, but rather specify a collection of rules that generate all the possibilities implicitly. Thus rather than listing individuals, we list types of individuals (such as {\small\tt User}); rather than listing actions, we specify a signature for each action type, such as {\tt\small login (username: String, password: String): (user: User)}; likewise we specify the type of each fact, such as {\tt\small registered (User)} and {\tt\small username(User, String)}. Finally, we specify a rule for each action type that says which facts it requires and which it ensures. All possible behaviors then follow inductively from these rules.\footnote{Initial conditions are not generally needed, because no individuals exist at the start and all facts are therefore false.}

\subsection{Origins}
This simple structure of individuals, values, actions and facts is sufficient to define the behavior of even the most complex software systems. It brings together three classic threads in software engineering. From formal specification languages (such as Alloy, B, VDM, TLA and Z) it takes the idea of defining behavior in terms of actions\footnote{Our approach might better be viewed as a hybrid of these languages with Hoare's CSP \cite{hoare-csp-book}, or as aligned with Event-B. In CSP, actions can be observed and their names matter. The same is true in Event-B. In other specification languages, in contrast, actions are often not treated as explicit phenomena. In Z, for example, actions are represented by named schemas, but the names are not given semantic significance, and only an informal convention interprets certain schemas as observable actions. Ironically, even in TLA, which stands for ``Temporal Logic of Actions'' there are no actions per se, only state transitions; the names of actions are irrelevant, and two actions that produce the same state transition are indistinguishable. Alloy has no hardwired ontology of phenomena, allowing modelers to decide for themselves whether actions are significant. A common pattern reifies actions as an explicit set of objects, which can then be shown in visualizations of executions and referred to in properties \cite{jackson-software-abstractions-book}. (Put differently, in state machine terms, our approach, in common with B, CSP, and Alloy with the above pattern, and unlike TLA and Z, defines behavior with a transition graph in which transitions are \textit{labeled}.) } with pre- and post-conditions over states. From early work on object orientation \cite{Abrial1974} it takes the idea of persistent individuals. And from entity-relationship modeling \cite{chen1976entity} it takes the idea of representing state as a collection of relations that associate individuals and that map individuals to values. Relationship types become relations; thus {\tt\small username(User, String)}, for example, would be a relation called {\tt\small username} from {\tt\small User} to {\tt\small String}.

\subsection{Introducing Concepts}
The meaning of even a small software system may involve a dozen types of individuals, and a hundred actions. Clearly some additional structuring is needed. Our approach is to use concepts \cite{jackson2021essence} as the organizing mechanism. Each concept corresponds to a functional concern or responsibility. The actions are partitioned among the concepts, and their associated facts follow them. In contrast, individuals need not be assigned to concepts; the same type of individual can participate in multiple concepts.

For example, the concepts of a social media application might include Authenticating, Posting, Commenting, Friending, Notifying and so on. The actions and facts described earlier would belong to the Authenticating concept. The Posting concept might have actions such as {\tt\small createPost (author: User, content: String): (post: Post)} and facts such as {\tt\small content (Post, String)}.\footnote{Don't be confused by the use of the same name for an action parameter and a relation:  in {\tt\small createPost (author: User, content: String): (post: Post)}, {\tt\small content} is the name of the second parameter of the action; in {\tt\small content (Post, String)}, it is the name of the relation that maps posts to strings.} Note that the {\tt\small User} type is not confined to any concept. Concepts often associate different properties with the same individuals. Thus a fact (aka relation) of the Authenticating concept maps a user to a username, and a separate concept, say UserDisplaying, could map a user to a display name.

\subsection{Coordinating Actions} \label{coordinating-actions}
One final detail remains. Actions between concepts must often be coordinated. For example, we might want to say that when a comment is made on a post, the author of the post is notified. This requires a coordination amongst three concepts: the Posting concept, which maintains in its state the facts about authorship of posts; the Commenting concept, which provides the action by which a comment is created and associated with a post; and the Notifying concept, which provides the action in which a message is sent to a user.

One might be tempted to express such coordinations as calls between actions. One could embed in the action that creates a new comment a call to a query that provides the author of the associated post, and then a call to the notifying action that performs the notification. But such an approach would complicate the single-step specification of actions, and, worse, it would couple the concepts together by introducing references between them.

Instead, therefore, we use the notion of concept \textit{synchronizations} or \textit{reactions}, which are declarative rules that sit outside the concepts, acting as mediators between them. For the example above we might have:

\begin{alltt}\small
\textbf{when:}
  Commenting.newComment (post, text): (comment)
\textbf{where:}
  Posting.author (post, author)
\textbf{then:}
  Notifying.notify (user: author, item: comment)    
\end{alltt}

The synchronization creates causal links between actions, binding the arguments of invoked actions to the results of earlier actions and queries. In this case, the user to be notified is the author of the post to which the comment was added, and the item conveyed in the notification is the comment itself.

\section{Applications}

\subsection{Meaningful UX}\label{usability-section}

Software is usable when its users can predict what their actions will do. The ontology in Section~\ref{structure-section} provides a language with which designers, users, critics, and other stakeholders can articulate precise, substantive accounts and analyses of
experiential phenomena in software. This novel expressivity has application to usability design, critique and amelioration of hostile and harmful design, end-user programming, and the broader social construction of meaning in software.

\subsubsection{Diagnosing Failures of Meaning}

When software fails to make its meaning clear, users build their own. \textit{Folk theories} are accounts of what the system is doing that fit the available evidence well enough to predict what to do next \cite{eslami-folk-theories, devito-folk-theories}. They are often the only resource a user has, and they keep working until they do not. When they fail, the failure is hard to report: the user knows something has gone wrong but lacks the words to say what.

Everyday file system actions in cloud-era software are full of such cases. Renaming a file may rename it for everyone with whom it is shared, or only for the current user, depending on whether the rename is a fact about the file or about the user's view of it\footnote{For an analysis of this issue in Dropbox, see \cite{jackson2021essence}.}. Sharing may mean granting access, sending a copy, or creating a link, each with its own rules for revocation, for the fate of the original when the share is removed, and for the recipient's visibility into the act. Deleting may mean removal from the current view, transfer to a trash whose retention policy is opaque, or irrevocable destruction; nothing on the screen distinguishes them reliably \cite{whittaker-files, voida-cloud-file-sharing}. The user is asked to act through one vocabulary---rename, share, delete---that the software implements via several concepts at once.

Branches in \texttt{git} are a developer-facing instance. The terminology and the available actions invite the reading that a branch is a context one can suspend and resume. But branching, as implemented, is not context switching: switching branches updates files in the working tree, and uncommitted work that would be overwritten must first be committed or stashed. The developer is therefore obliged to operate a second concept (\texttt{stash}) whose name does not suggest its use for context management. An official mechanism for this is \texttt{worktree}, which separates concerns of working state from concerns of branching \cite{git-worktree-docs}. That the fix sits two pages into the documentation, and that the workaround continues to be taught to new developers, shows how reluctantly such conceptual mistakes are unmade.

The diagnosis in both cases is the same: the action a user wants to perform does not exist as an action of the system. Distinct concepts have been conflated under a single gesture, and the work of design is to pull them apart. Worktrees are the right kind of fix for branches; similar pulling-apart could be done for many of the file system operations above. The precondition is a vocabulary in which the diagnosis can be presented to the people in a position to act on it.

\subsubsection{Misleading Meaning and Dark Concepts}

The cases above are conceptual mistakes. Worse are cases in which the misalignment is intended. The literature on \textit{dark patterns} catalogs interfaces engineered to mislead users into outcomes they did not intend, but it tends to stay at the level of the screen: a button too small, a confirmation too aggressive, a cancellation path too deeply buried \cite{brignull-dark-patterns, mathur-dark-patterns}. These are usually just symptoms of a deeper malaise: a mismatch between the action the user thinks they are taking and the action the system records.

Facebook's reactions, raised at the start of this paper, illustrate the point. A user who clicks the angry button is invited by the interface to perform an act of expression: this post made me angry. The system, for some years, treated the click as a strong signal of engagement, weighting it more heavily than a like by a documented factor of five \cite{merrill-oremus-angry}. The act the user performed and the act the system recorded were not the same act. From the screen alone the discrepancy was undetectable; what changed was the shape of the feed.

Concept design offers a way of naming the offense. ``Dark pattern'' leaves the kind of misalignment open; calling angry-as-amplification the implementation of one concept (engagement weighting) under the guise of another (sentiment expression) locates the offense at the level of meaning. The same kind of naming applies elsewhere: to subscription cancellation flows routed through pages of dissuasion, to free trials whose registration is a single action and whose revocation is governed by no comparable one, to permissions dialogs that grant in one step and revoke only in many \cite{ftc-click-to-cancel}. The asymmetry, in each case, is a property of the concept rather than of the screen, and naming it there is what makes it open to analysis and to regulation.

\subsubsection{Enshittification: A Standard Trajectory}

Cory Doctorow's account of \textit{enshittification} describes the trajectory along which platforms degrade over time: first they treat their users well, then they extract value from those users to lock in business partners, and finally they extract value from the partners until little remains \cite{doctorow-enshittification}. The mechanism, in concept terms, is a steady reinterpretation of the concepts the platform was originally understood to provide. \textit{Searching} on a marketplace once retrieved the items most relevant to the query; on a mature platform it retrieves the items the seller has paid most to surface. \textit{Following} once meant subscribing to a creator's posts; on the same platform later in its life it means receiving them in an algorithmically thinned trickle. The user's action does not change; the concept behind it does. Enshittification, in this reading, is not a sequence of new dark patterns layered on a stable system; it is the gradual substitution of one concept for another behind an unchanged interface, and it can be sustained for as long as users keep acting on the old reading. A vocabulary in which both readings have names is what allows the substitution to be noticed.

Winograd and Flores argued long ago that design is ontological: it constructs the world of actions and entities through which the user thinks about the software \cite{winograd-flores}. When designer and user interests diverge, the user's recourse is bounded by what the design admits as an action. Proposals to broaden that recourse---to let users rewrite or reinterpret the ontology, as in some end-user customization tools and federated social platforms---are, at root, proposals to redistribute authority over meaning \cite{graffiti, zignani-mastodon}.

\subsubsection{A Shared Language for Designers and Users}
 
Meaning operates on a second interface too: between the user and the team that builds the software. Users participate in design through bug reports, through user research, and sometimes through deeper involvement in design itself \cite{simonsen-robertson-participatory-design}. The vocabulary available to that participation bounds what it can achieve.

Take the user story, the canonical artifact of agile development: \textit{As a [user], I want [action] so that [benefit]} \cite{cohn-user-stories}. In practice the named action is a placeholder, a piece of prose the development team will negotiate into something implementable. The negotiation happens in private and its result is rarely traceable to the original story. When the project maintains a vocabulary of concepts, actions and reactions, the user story can use that vocabulary directly: \textit{as a poster, I want an action on \texttt{Posting} that schedules a post for a future time, with a reaction that publishes it when the time arrives}. The form is more demanding, but it leaves much less room for interpretation, and the discussion that follows can be about whether the design is the right one, rather than an attempt to clarify what design is actually being proposed.

Bug reports work the same way. ``This feels broken'' is hard to act on; ``\texttt{React.angry} seems to participate in two reactions, one feeding ranking and one notifying the author, and the two weight it differently'' is something a developer can investigate directly. A report of the second kind requires a user who has a clear understanding of the meaning of the software and its key elements. The information-needs literature has long observed that what developers want and what users supply rarely match \cite{ko-information-needs, bettenburg-bug-reports}. A shared concept vocabulary is one way to close part of that gap.

Participatory design aims to treat users as collaborators rather than informants \cite{muller-participatory-design}. Its standing difficulty is that users and designers do not speak the same language. Users describe what they experience; designers describe interfaces and data models; the designer's interpretation bridges the gap, with the asymmetries that brings. An explicit conceptual model offers something else: an artifact the user can read, push back on, and propose changes to in the same terms the designers use. Disagreements can then be pinned to the model rather than absorbed silently by the designer.

\subsubsection{End-User Programming}
End-user programming has long aspired to let ordinary users compose their software from pieces that match the activities they want to perform \cite{nardi-vision, ko-state-of-art}. In practice the available pieces---spreadsheet formulas, macro languages, IFTTT-style triggers, low-code blocks---tend to fall short in one of two ways. Either they expose too much (raw syntax, with no notion of what its parts \textit{mean} beyond their evaluation), or too little (rigid templates that do not compose). The user is left without units of meaning at the level at which they actually think.

Concepts can provide such units. Each is named, scoped to a single concern, and connected to other concepts only through declared reactions. A user who knows what \textit{Posting}, \textit{Commenting} and \textit{Notifying} are, and who can read a reaction such as ``when a comment is made, the author of the post is notified,'' can inspect and modify their software at the level at which they use it. How the rule is executed can be left to the implementation; whether the rule is the right one is a question they can answer for themselves.

Large language models are now widely promoted as the answer, on the premise that natural language has at last become an effective programming interface \cite{sarkar-llm-end-user}. We are skeptical. Such models generate syntax from prose, which is useful, but they do not generate a vocabulary in which the user can reason about what the software does. A user who asks a model to ``add a feature'' and receives a working patch is no better placed than before to say what the software now means. Concept design supplies the missing layer. With a conceptual model in hand, the user can request changes at the level of meaning---a new reaction, an additional fact, a new action of an existing concept---and read the result in the same terms.


\subsubsection{The Social Stakes of Meaning}
Meaning, in everything we have said so far, is something a designer makes and a user receives. But meaning is also contested. The same artifact, in the same form, can be read very differently by people whose positions and interests differ.

Ruha Benjamin's \textit{Race After Technology} traces how systems presented as technical---recidivism scores, hiring filters, facial recognition---encode and reproduce social inequities under a posture of neutrality \cite{benjamin-race-after-technology}. The disagreements that arise around such systems are usually not, in the end, about implementation. An engineer who asserts that an algorithm meets its benchmark and a sociologist who reports that its deployment harms a community are not disputing each other's claims. They are disagreeing about what the algorithm \textit{means}. The engineer reads it through its evaluation pipeline; the sociologist reads it through its consequences in the institutions where it is deployed. Each reading is internally consistent. Neither uses the other's language \cite{suchman-located-accountabilities, selbst-fairness-abstraction}.

A vocabulary of explicit meaning offers something small but useful here. The phenomena it makes explicit---actions, facts, and their participants---give technical and social analysis a common ground. To describe a risk-scoring system as an action that takes a person and produces a classification, and the classification as a fact recorded against that person and read by later actions of a court, is to describe it in a way that both engineer and sociologist can recognize. The structural questions about it---which facts persist, who can write them, who can read them, who can revise them---belong in a single conversation. A regulator's worry about the irrevocability of a classification and an engineer's worry about its calibration become two angles on the same artifact, asked in the same terms.

This does not automatically resolve disagreements, and we do not claim that explicit meaning is sufficient to eliminate contention, or to make software just in everyone's eyes. But many of the public misunderstandings around algorithmic systems have followed from the absence of a vocabulary in which both technical and social claims can be made. Supplying one is a prerequisite for a discussion that has, so far, been very hard to have.

\subsection{Building Code from Meaning}\label{code-gen-section}

The patterns of code today are structural: objects, functions, variables, types, and 
more complex 
idioms
that form the programmer's building blocks. Software development is then the practice of building a meaningful structure from these patterns that will serve a greater purpose than the mere assembly of parts. In many cases, the names found in code---despite having no formal semantic meaning by definition---are the most meaningful elements of software. Two functions otherwise identical in formal semantics and implementation, such as notifying mechanisms named accept() and acknowledge(), might have very different ramifications.

What if our patterns for code began from meaningful and intentional names? Applying concept design to code is the process of software development flipped on its head: instead of assigning meaning to structures, the goal is to assign structures to meaning. The patterns that implement a concept, such as the choice of data structure or control flow, are secondary to its names (concept, actions, parameters). The guiding principle for assigning the phenomena of individuals, values, actions, and facts to code is simplicity, and the pattern that most succinctly captures the phenomena is preferred.

Considering the phenomena in turn:

\begin{itemize}
    \item \textbf{Individuals $\rightarrow$ UUIDs.} An individual is minimally represented as a unique identifier. In the same way that individual people are free to participate in an open and wide variety of actions, so too must the code enable the same flexibility (i.e. of passing around a unique identifier). An object with a fixed set of methods would misrepresent the capabilities of an individual and would be a brittle abstraction.
    \item \textbf{Values $\rightarrow$ value-objects.} Almost every language today has a notion of (mostly) shared primitive values, such as integers, characters, and strings. Together with collection types (sets, lists, maps), these value-objects are treated as interpretable by their structure and content alone.
    \item \textbf{Actions $\rightarrow$ functions over maps.} Actions involve both individuals and values, and have an input and output with varying arity. A function that takes a map and outputs a map, where the keys are the named parameters of the action, satisfies both requirements while allowing for a uniform representation of actions (e.g., for logging and synchronization).
    \item \textbf{Facts $\rightarrow$ relations.} A relation can be implemented in many ways: canonically in relational databases, normalized into a triple-store, or rendered into trees in a document store with manual management of invariants. 
\end{itemize}

Building on these choices, the larger idiom of concept design can be implemented in TypeScript as follows:

\begin{itemize}
    \item \textbf{Concept $\rightarrow$ singleton class.} Instead of representing individual users as objects with behavior that may need to be migrated over time, concepts are mapped one-to-one with a stable single instance of a class concerned only with one concern, such as Authenticating. This class forms a namespace and isolation boundary for the lifecycle of the concept, consolidating all necessary implementation details in one location.
    \item \textbf{State $\rightarrow$ attributes.} State is held internal to the concept class, and in practice may be persisted to external databases. Regardless, either the full state or a reference to the state (e.g. database connection) is held as attribute(s) that can relate individuals (UUIDs) and values.  
    \item \textbf{Action $\rightarrow$ method.} Concept actions are methods that take in a JSON map and output a JSON map, corresponding to its parameters. In the body of the method, any necessary computations and side-effects are performed, such as updates to the state.
\end{itemize}

These choices prescribe very few restrictions on implementing concept design in a language like TypeScript. Notably, when implementing concepts, there are no requirements for extending a custom class, importing a framework, or using any DSL. A vanilla TypeScript class that adheres to the convention of all actions taking and returning a JSON map is sufficient structure for a concept.

\subsubsection{Organizing Code Around Meaning}

As described in section \ref{coordinating-actions}, synchronizations are the mechanism for composing concepts. They provide a minimal DSL for establishing the mechanism by which all actions are fired: \textbf{when} a matching action occurs, \textbf{where} certain conditions on state are held, \textbf{then} the following actions are called. To implement this within code, a reference TypeScript framework is built entirely as a tiny, zero-dependency engine that parses and processes a synchronization domain-specific language (DSL). The DSL is represented fully as compliant TypeScript with type-checking, and features a single ``action(...)'' helper to help write TypeScript that corresponds directly to informal specifications of synchronization. 


This engine enables the following organization of code in a repository:
\begin{itemize}
    \item \verb|src/concepts/| --- Each concept class is stored within its own folder (along with any test code) as a single TypeScript file. 
    \item \verb|src/syncs/| --- All synchronizations can be declared here as fully independent blocks of code and saved in any arbitrary nested folder structure with the file naming convention of \verb|name.sync.ts|.
    \item \verb|src/engine/| --- Provided engine code that developers do not need to touch.
    \item \verb|src/main.ts| --- Provided entrypoint script that discovers and registers all concepts (using proxies to provide reactivity over arbitrary TypeScript classes) and wires all declared synchronizations.   
\end{itemize}

Developers using this framework only ever touch the \verb|concepts/| and \verb|syncs/| folders in the source code. Additionally, existing concepts may be dropped in without any modification or configuration, and made useful by adding synchronizations that refer to the added concepts. The act of developing the source code is itself extending a meaningful vocabulary: new concepts to describe lifecycles of behavior that are relevant, and new synchronizations to describe how the application should specifically act. 

\subsubsection{Meaningful Modularity}

The key property of this architecture is that the code structure maps directly to units of meaning: a developer only ever has to write concepts as independent TypeScript classes, and declare how they compose as synchronizations as individual granular rules. There are no additional configurations to manage, nor abstractions that do not map to meaningful notions: layers such as middleware, API servers, dependency injection, etc. are not necessary. Instead, any essential functionality can be encapsulated as a more meaningful concept. In a web application, for example, receiving and responding to HTTP requests can be encapsulated in a Requesting concept.

This idea of the Requesting concept demonstrates how concept design can help simplify the story of developing a web application into a more meaningful story. Instead of stacking the entire application logic on top of a laundry list of patterns that would confuse an average end-user (API, router, handler functions, middleware, etc.), the Requesting concept has two actions: \textbf{request} and \textbf{respond}. The \textit{path} parameter of the \textbf{request} action specifies the URL of the request, and enables synchronizations to fulfill the role of a router by pattern matching. When a user makes a particular request, the proxied Requesting concept records the action, path, and any additional parameters, which triggers any declared synchronizations. Another granular synchronization, which could have a \textbf{when} clause awaiting completion of a task, would then fire the \textbf{respond} action of the Requesting concept to respond to the request.

What differentiates this approach from existing web development frameworks is not modularity of \textit{structure}---many of these frameworks were invented precisely to factor out code into separate parts---but rather modularity of \textit{meaning}. By devising a single concept to handle the notion of requesting a resource, concept design ensures that both users and developers need only understand one concept to understand the lifecycle of a request. If they wish to discover how that request intersects with other concerns, such as actual deletion of posts or notification of friends, they can identify and analyze all synchronizations involving exactly that set of concepts.

\subsubsection{Legibility}

The affordances provided by meaningful modularity establish a clarity that affects both usability and accountability. This property---of being able to map meaning in the world to and from the code---can be understood as the \textbf{legibility} of software. These ideas draw on previous work \cite{meng2025legible} about legible software, and argue for understanding and evaluating software in ways that functionally change what it means to its users. Updating scattered authentication logic into middleware layers is not a legible change for a user who wants to understand what they can do while logged in. On the other hand, a deterministic list of all reactions involving {\tt\small Sessioning.loggedIn} in the \textbf{where} clause is both the code that implements the logic, as well as a readable and reliable contract for what the system does.

The legibility of a project involves both the code at definition (compile-time) and during operation (run-time). A key aspect of the concept design framework is the action log, which provides a linear history of every action of every concept. This log is the source of truth: a list of behaviors that occurred in the past is an immutable account, and the state of concepts can be reconstructed entirely from the log. One key aspect is that both action invocations (the request, with its input parameters) and completions (what actually happens, and whether it errors or not) are committed to the log. Even for non-deterministic actions, the state can be recovered deterministically from the recorded completions.

The benefits of the action log for legibility include:
\begin{itemize}
    \item \textbf{Accountability.} Every action of the system is not only recorded in the log, but also given \textit{provenance edges} to the actions that led to it. These edges are labeled by the synchronization that caused it, which is itself an addressable element in the code and uniquely named in the specification. This links all actions (which must be the result of some \textbf{then} clause) directly to the declared action patterns in the \textbf{when} clause, establishing the invariant that each and every action is only committed to the log by way of a synchronization.
    \item \textbf{Reliability.} Actions experienced by users are strongly consistent, and users may reliably expect the system to react to actions they observe. Synchronizations are only fired after actions are committed to the log, and the separation between invocations and completions allow the system to survive crashes during processing. The full reification of the execution history itself also enables replay of actions from subsets of history, deterministic testing, and what-if scenario branching. 
    \item \textbf{Usability.} While the system shares properties (such as time-travel replay) with event-based patterns like event sourcing, one key distinction is that the actions are written with a granular and usable schema---held in common by all parties sharing the associated concept---that provides a meaningful account on its own. Unlike events, which tend to grow increasingly illegible in systems where downstream processors require more and more metadata, action records are a stable and readable surface. As users engage with a system, each of its actions (which define the entire state) must map closely to their understanding of how they should use it and how others have acted. 
    
\end{itemize}

\subsubsection{Experience from Teaching}
Concept design was taught in a senior-level software design course at our institution in the fall of 2025, using the lightweight TypeScript framework. At the same time, the course promoted and taught how to collaborate with LLMs to build software using a specification-first approach and principled context management. Over the course of the semester, roughly 80 individual projects and 20 team projects of significant scope were completed using the approach. Each project was structured as a set of reusable concepts, and a set of application-specific synchronizations.

Students were provided with a sample repository implementing a fully functioning file sharing application, derived from a set of commonly useful concepts such as Sessioning, Authenticating, and FileUploading. They were able to drop not just the concepts, but a default set of common synchronizations (such as with Requesting, establishing a shared set of API routes for logging in) into their repositories to implement a working authentication system. Whenever a new route was needed, they did not need to modify the Requesting concept, but could instead add a new synchronization that matches on the desired path.

For new concepts and for implementing synchronizations in general, students leveraged LLMs extensively, and were provided with a set of background documentation that simultaneously served as a tutorial on concept design, as well as context for LLMs. The full flow of building concept-driven systems with LLMs can be broken down as follows:

\begin{enumerate}
    \item \textbf{Problem framing.} Describe what the system should achieve and what problems it solves as a set of text files (e.g. markdown).
    \item \textbf{Designing concepts and behavior.} Work with an LLM to design the conceptual breakdown of the problem into individual concept specifications, and declaring the behavior between concepts as synchronizations. Students were able to describe and reason about concepts in prose, and LLMs were capable of devising fully-compliant concept specifications given the background documentation.
    \item \textbf{Generating concepts.} A single LLM call generates the implementation from the specification for each concept: given the simplicity of concepts, this was observed in practice to be highly reliable. Concepts may be tested on their own, and are strongly isolated from other concepts and from any custom framework code.
    \item \textbf{Generating synchronizations.} LLM calls with synchronization specifications as input provide synchronization code that is incorporated into a running system by virtue of placing it in the \verb|syncs/| folder. Each synchronization corresponds to one observable behavior, and can be developed in isolation, and tested against arbitrary histories in the action log. 
\end{enumerate}

As no concept can import another, LLM generation could proceed both in parallel and with a much smaller set of context: only the shared background documents were necessary, and no concept needed to understand either other concepts, or synchronizations. For synchronizations, only the declared concepts would need to be referenced. The students were provided with a lightweight context management tool to make each of the steps above a single CLI command pointing at a markdown file of their choice (e.g. the concept spec to generate its implementation).

Many students were tempted to allow commercial LLM coding agents to run on their repositories. Due to the unrestrained nature of these agents, significant difficulties arose:
\begin{itemize}
    \item \textbf{Cost.} Agents run into usage limits quickly, and a number of students, who had delegated implementation work aggressively, reported being rate-limited on their ability to work on their assignments. A number of our experiments comparing the cost of implementing a single concept measured at less than \$0.09 for properly managed context, as opposed to more than \$3 for a commercial agent.
    \item \textbf{Breaking the build.} Agents sometimes decided to modify engine code as a shortcut to implementing behavior that belonged within a concept or synchronization. 
    This behavior was alarming but easily detectable given the modularity discipline that changes should only affect the concepts and syncs folders. No example of a modification to engine code was found to be legitimate, and every case could be refactored more simply in a concept.
    \item \textbf{Agentic debt.} Beyond accruing technical debt, students who made heavy use of commercial agentic tools exhibited a loss of agency over their own repository, their understanding, and their strategy for development. This can be thought of more broadly as a form of debt that describes the compounding price of delegating decision-making to agents without staying in the loop and driving meaningful exploration.
\end{itemize}

At the end of the semester, a common thread found within student reflections was a heightened appreciation of the importance of proper modularity (``not just an abstract principle'') and how their experiences engaging with concept design surfaced that idea in clear and obvious ways. Students also reported how the concept design discipline aided with LLM code generation, enabling more possibilities (``a lot more creative freedom'').

\subsection{Agentic Code of Conduct}\label{agent-section}

Of all the settings in which meaning matters, the design of autonomous
agents may be the most demanding. An agent that has been granted
authority over a user's files, email, calendar or experimental codebase
acts in the user's absence. What it did, why it did so, and whether it
had the right to do so must be answered from a record largely produced
by the agent itself. Direct observation, which mediates most software
use, is no longer available.

The conventional artifacts of agentic systems are poorly suited to
this. An agent is typically specified by a prompt and a set of tools,
and what it does is recorded as a sequence of tool calls and
intermediate messages \cite{yao-react, wu-autogen, wang-voyager}. The
prompt is discursive prose, written for a model to interpret; the
trace is detailed but semantically thin. Neither carries the meaning
of what the agent was doing in terms a stakeholder can recognize. Was
a particular commit made in service of a hypothesis formed in advance,
or rationalized after the fact? Did a metric improve for the predicted
reason, or by coincidence? Was an email sent because the user
endorsed it, or because the agent inferred a wish from indirect
evidence? These are not questions about \textit{which} tool was called. They
are questions about \textit{meaning}.

What is needed---and what we believe our approach naturally
provides---is a structure that names the activities the agent
performs, states the conditions under which they are permitted, and
records observations in the same vocabulary in which those activities
are described. We call such a structure a \textit{code of conduct} for
the agent\footnote{The runtime mechanisms by which such codes can be
enforced, and a more detailed account of the experiments mentioned
below, are currently under review in a submission to a different
venue.}.

\subsubsection{Concepts and Reactions}

Consider an agent engaged in autonomous research, of the kind explored in recent demonstrations and systems \cite{karpathy-autoresearch, lu-ai-scientist}. Over the course of a run, it forms hypotheses, edits code, plans and runs experiments, records measurements, and communicates results. Each of these is a concept in the sense of Section \ref{structure-section}, with its own actions and facts. \textit{Hypothesizing} owns the actions by which a hypothesis is formed and the predictions it carries. \textit{Editing} owns the actions by which changes to the codebase are proposed, applied to a working copy, and committed to the repository. \textit{Experimenting} owns the planning and execution of experiments. \textit{Evaluating} owns the recording of measurements, with the discipline that recorded values come from execution rather than from the agent's own description of what it observed.

The synchronizations among these concepts are written as reactions of the kind already introduced. The simplest useful reaction for an experimental coding agent is that a change may be committed only when bound to a hypothesis formed in advance:


\begin{alltt}
\small\textbf{when:}
  Editing.apply (change)
\textbf{where:}
  Hypothesizing.form (hypothesis, supports: change)
\textbf{then:}
  Editing.commit (change)
\end{alltt}

The agent can apply a change to a working copy at any time to inspect its effect, but only commits it when the change is bound to a hypothesis. Without such a binding, the apply has no further effect. Other reactions extend the pattern: a commit can require a planned experiment, whose result then enters \textit{Evaluating} and is checked against the hypothesis's prediction. In each case the step is named, and what permits it is stated in terms of other named steps.

A code of conduct is a small document in which the concepts and the reactions among them are written down. It is short enough to be read directly and precise enough that, for any action the agent takes, the clause that caused it can be identified. Because the code and the execution trace use the same vocabulary, a reviewer can read them together: the trace reports what happened, the code states what would have made it legitimate, and the alignment or misalignment between the two becomes part of the record.

This gives the trace a status it does not have when it is merely a log of tool calls. Predictions, in particular, play a prominent role. A hypothesis is admitted only if it carries a prediction; an outcome is admitted only if it compares the observation against the prediction it was meant to adjudicate. A metric that improves while contradicting its predicted mechanism then appears in the trace not as a silent success but as an explicit disagreement between what was expected and what occurred.

\subsubsection{Properties of the Governed Trace}

The value of writing such a code of conduct reaches beyond governance, and its
benefits follow from a single structural choice: the trace
consists of typed actions over typed facts rather than free-form prose
interleaved with opaque tool calls.

The first benefit has already been stated. Because code and trace share
a vocabulary, any action in the trace can be pinned to the clause that
authorized it, or seen immediately to lack one. The remainder are less
obvious.

Long agentic runs commonly fail in ways loosely attributed to memory:
an agent revisits a path it has already explored, repeats a question
it has already answered, or proposes a fix it has already tried. The
prevailing approaches treat this as a problem of context capacity and
propose retrieval-based remedies that summarize prior transcripts or
page them in on demand \cite{packer-memgpt, park-generative-agents}.
A meaning-governed framing suggests a complementary diagnosis. The
difficulty is not only that prior material has scrolled out of view,
but that even when it remains in view, prose is hard to search by
intent. A trace organized around typed actions and the facts they
update is queryable in the same vocabulary in which it was produced:
an agent resuming work can ask which hypotheses it formed, which
experiments tested them, and what those experiments found, rather than
reading its own narrative back and reconstructing the answers. The
same property persists across runs, giving the trace a second role as
durable memory: findings, decisions and predictions are stored in a
form that admits direct query along the structure of the work itself,
not as transcript fragments to be retrieved by similarity.

As a further benefit, traces become replayable. Given any
prefix of a real run, a different model---or the same model under
different conditions---can be asked what action it would take next
under the same code. Because actions are typed, the two candidate
continuations are commensurate, and the comparison meaningful. We see
this as a route to a kind of evaluation that does not require
benchmarks constructed in advance: traces recorded under codes of
conduct become a corpus of decision points at which models can be
measured against one another, on the work the agents are deployed to
do rather than on tasks designed as proxies for it.

A final benefit applies when more than one agent is involved. In existing multi-agent frameworks such as AutoGen \cite{wu-autogen}, agents coordinate by exchanging messages: each agent receives messages from others and decides how to respond, and the joint behavior of the system is the trajectory of the resulting conversation. Under a shared code of conduct, agents are instead governed by a single set of concepts and reactions, and a single trace records their joint activity in a shared vocabulary. An agent passing a hypothesis to another for testing does so through an action in \textit{Hypothesizing} whose effect, by synchronization, is an action in \textit{Experimenting}. The handoff is part of the structure, not an artifact of how the agents happen to communicate.

\subsubsection{Experience}

We have applied codes of conduct of this form to three autonomous research loops. The first was a small language-model training agent in the style of recent autoresearch demonstrations \cite{karpathy-autoresearch}: the agent edited a training script, ran training, measured validation loss and decided which changes to keep. The second applied the same discipline to an exploratory optimization problem, in which the agent applied changes intended to reduce the runtime of a numerical simulation. The third concerned semiconductor TCAD, the use of computer simulation to predict the behavior of a device before fabrication. Here several agents worked in parallel under a shared code of conduct, each pursuing its own line of inquiry, with a supervising agent auditing recent activity and recording durable findings.

In each case the runs produced real results. The training agent found changes that reduced validation loss; the optimization agent achieved measured speedups in the simulation's runtime; the TCAD agents moved the simulated device closer to its target specification\footnote{The PhD student in a micro-technologies lab who applied our TCAD workflow reported progress of weeks worth in just a few hours.}. One might worry that a code of conduct trades results for legibility, with an agent confined to named, attributable steps doing less than one free to range over tools and prose. We did not find this. Legibility came alongside the results, not at their expense.

The runs differed in shape, but the same pattern recurred. Where conventional artifacts---a commit history, a metrics file, a chat transcript---would have collapsed the work into a single number or an unstructured narrative, the code-governed trace preserved the structure that made the work intelligible. In the training run, it revealed an agent revisiting the same hyperparameters from different angles without registering that it had done so before, a kind of search without memory, most likely caused by drift beyond the agent's effective context. In the optimization run, it preserved an empirically successful change whose stated mechanism could not account for the measured speedup, separating a real result from a failed explanation. In the TCAD run, where the work branched across agents rather than proceeding serially, the same discipline applied once the concepts were adjusted: actions for branching from a prior run, for auditing a recent window of activity, and for recording durable findings replaced the serial vocabulary, and the trace read as a coherent campaign rather than as an accumulation of unrelated branches.

We draw only modest conclusions from these cases. They are small, and the codes of conduct used were specific to their domains. But they suggest that the same basic ideas---naming activities as concepts, stating the reactions among them as synchronizations, and recording a trace in a common vocabulary---can govern several different kinds of agentic work. When the shape of the work changed, the underlying machinery did not; only the concepts did. And in each case, the trace acquired substantive meaning in exactly the sense this paper has argued for: a shared vocabulary, grounded in observable phenomena, in which what the software does become inspectable by those on whose behalf it acts.

\section{Future Prospects}

As many programmers question their role in a new era in which most code is written by AI agents, a new possibility emerges: that their value will come from creating and shaping meaning.

Thus making software meaningful is not an ex post facto effort to retrofit meaning to software but a new way to derive the production of software from its meaning. Perhaps this will also alleviate the sense that programming work is no longer valuable, by ensuring that not only the software, but also the making, is meaningful. And perhaps it will encourage also the development of more meaningful software, that is more attuned to the needs and understanding of its users.

While this proposal is aligned with current technology trends (decomposition into microservices, event-driven architectures, specification-driven development, etc.), it is nevertheless counter-cultural. Programmers see their task as writing code; UX designers produce wireframes and user journeys; database engineers define schemas; product managers create feature roadmaps; and so on. 

If software is to be made meaningful, the most pressing task must be to create a shared meaning that guides and shapes all other activities. Programmers may complain (unjustifiably) that this takes them away from technical to social and psychological concerns, and UX designers that naming and defining actions feels more like API specification than the human-centered work they prefer to do.

These perceptions are precisely what has led us to the current situation in which gaping gulfs emerge between software and users, between designers and programmers, and between real needs and delivered features. Only by committing to an explicit and shared understanding of meaning will these problems be decisively overcome.

\bibliographystyle{ACM-Reference-Format}


\end{document}